  \documentclass[aps, amsmath,amssymb,floatfix]{revtex4}

\usepackage{dcolumn, graphicx,color, bm} 
\usepackage{ulem} 

\begin{document}
\title{Inverse cascades in turbulence and the case of rotating flows} 

\author{A. Pouquet$^1$, A. Sen$^1$, D. Rosenberg$^1$, P.D. Mininni$^{1,2}$,  and J. Baerenzung$^3$}
\affiliation{
$^1$Computational and Information Systems Laboratory, NCAR, 
         P.O. Box 3000, Boulder CO 80307, USA. \\
$^2$ Departamento de F\'\i sica, Facultad de Ciencias Exactas y Naturales, UBA \& 
         IFIBA,CONICET, Ciudad Universitaria, 1428 Buenos Aires, Argentina. \\
$^3$ Interdisciplinary Center for Dynamics of Complex System, D-14476 Postdam, 
         Germany.}
\date{\today}
\begin{abstract}
We first summarize briefly several properties concerning the dynamics of two-dimensional (2D) turbulence, with an emphasis on the inverse cascade of energy to the largest accessible scale of the system. In order to study a similar phenomenon in three-dimensional (3D) turbulence undergoing strong solid-body rotation, we test a previously developed Large Eddy Simulation (LES) model against a high-resolution direct numerical simulation of rotating turbulence on a grid of $3072^3$ points. We then describe new numerical results on the inverse energy cascade in rotating flows using this LES model and contrast the case of 2D versus 3D forcing, as well as non-helical forcing (i.e., with weak overall alignment between velocity and vorticity) versus the fully helical Beltrami case, both for deterministic and random forcing. The different scaling of the inverse energy cascade can be attributed to the dimensionality of the forcing, with, in general, either a $k_{\perp}^{-3}$ or a $k_{\perp}^{-5/3}$ energy spectrum of slow modes at large scales, perpendicular referring to the direction of rotation. We finally invoke the role of shear in the case of a strongly anisotropic deterministic forcing, using the so-called ABC flow. 
\end{abstract}
\maketitle
  
\section{Introduction}

The present renewed interest in two-dimensional (2D) turbulence is due to several factors. On the one hand, the increased power of computers allows for modeling flows with higher Reynolds numbers, and thus for a more accurate determination of the statistics and characteristics of such flows. Moreover, the observation in the atmospheric boundary layer of an energy spectrum with a  $k^{-3}$ law followed at smaller scales by a $k^{-5/3}$ law \cite{nastrom_84} has led to some controversy since the opposite case was thought to happen (see below): the atmosphere is viewed as quasi-2D at large scales as recent analysis of data with modern velocimetry techniques shows rather unambiguously \cite{tellus}, and thus it is expected to undergo an inverse cascade of energy to large scales.

The inverse energy cascade in 2D flows has first been observed  numerically \cite{hossain_1983}. It was later confirmed experimentally in \cite{sommeria_86} using an electrically driven flow in a thin layer of mercury in a square box, and in numerous configurations since these pioneering papers \cite{tabeling, boff_rev}. There are other 2D systems undergoing inverse cascades, when incorporating more physical phenomena, such as beta-plane turbulence or multi-layer quasi-geostrophic (QG) flows \cite{vallgren_2010}, as well as when coupling the velocity dynamics to that of magnetic induction (although in the case of conducting fluids, it is not the energy that populates the scales larger than the forcing scale). For very strong rotation (at fixed Reynolds number) it is known that three-dimensional (3D) flows tend to become 2D and thus to remain non-singular in the limit of zero viscosity \cite{babin}; indeed, an inverse cascade of energy was observed in \cite{smith_96} for such flows, and the co-existence of an inverse cascade of energy to large scales and of a direct cascade of energy (and of helicity) to small scales was studied in \cite{mininni_TG, mininni_1536a} (see \cite{phil_trans} for a recent review), with an eventual recovery of isotropy and Kolmogorov scaling at small scales \cite{mininni_3072}. 

The inverse cascade of energy in 2D turbulence was postulated by Onsager \cite{onsager, eyink} when studying the interactions of an ensemble of point vortices, and by Kraichnan \cite{rhk_67} using statistical equilibria of a finite number of degrees of freedom in the ideal (non-dissipative) case. This cascade is characterized by a transfer of energy with a constant flux, up to the largest scale accessible to the system; it is attributed to  the dual constraint of total energy and squared vorticity (enstrophy) conservation for inviscid 2D flows. The review article written on 2D turbulence by Kraichnan and Montgomery \cite{rhk_montgo_80} concentrated on the theoretical issues, including in the case of coupling to a magnetic field. It stated that  the inverse cascade for 2D fluids should follow a $E(k)\sim k^{-5/3}$ spectrum (similar to the Kolmogorov  spectrum in the direct cascade of energy in three dimensions), together with a $k^{-3}$ law for the energy in the direct enstrophy cascade to small scales \cite{leith, batchelor}. Steeper spectra at small scale have been observed \cite{borue_93, verma_11a}, in particular in early numerical simulations at low resolutions; sometimes they are related to the dominance in such computations of strong coherent vortices.

In physical space, the inverse cascade corresponds to the formation of large-scale vortices (jets \cite{danilov_01b}, or bars \cite{yin_03}, can also be found). These structures have been observed for quite a while in numerical simulations \cite{hossain_1983} and in the laboratory \cite{sommeria_86}. They can be viewed as being due to a negative eddy viscosity arising from small-scale eddies \cite{rhk_76}, or as a non-local interaction between a small-scale vortex embedded in a large-scale strain \cite{chen_2006}. However, it has been known for a long time as well that the large-scale spectra can be steeper than $k^{-5/3}$, especially close to the gravest allowable mode where the energy condenses. In this case, the long-time energy accumulation at the largest scale is viewed as providing a source of energy for a downward cascade to smaller scales through filamentation of the large scale vortex.  The large scale flow can, in turn, decrease the level of turbulence \cite{shats},
as also found in the context of plasma flows \cite{biglari}. When Ekman friction is present, steeper spectra in inverse cascades can also be interpreted as being caused by a wavenumber-dependent energy flux arising from the friction, similar to a phenomenon already documented for magnetohydrodynamics flows in the quasi-static approximation relevant at low magnetic Reynolds numbers \cite{verma_11b}.

These so-called {\it condensates }\ undergo random reversals, more temporally sparse 
 when large-scale friction is diminished, the coupling to high-frequency modes providing the random noise that can trigger the transition from one large-scale quasi-steady state to another one \cite{dmitruk2011}; this phenomenon is observed in Rayleigh-B\'enard convection, in 2D Navier-Stokes turbulence in boxes with close to unity aspect ratio \cite{bouchet}, in fluid experiments \cite{cler}, in dynamo experiments \cite{pinton_reversal}, and observed as well in the reversals of the Kuroshio oceanographic current \cite{kuroshio} and of the magnetic field of the Earth \cite{valet}.

The {turbulence} statistics in the inverse cascade is Gaussian until the formation of the condensate, when the spectrum steepens \cite{smith_yakhot_94}: the inverse cascade is known to be self-similar, with a linear variation with order of scaling exponents of structure functions of the velocity field. However, Smith and Yakhot  \cite{smith_yakhot_94} noted that the condensate is responsible for intermittency  at large scale, viewed as a finite size effect due to the fact that the inverse cascade of energy has reached the gravest mode, and the boundary of the system where (Ekman) friction can play a role as well. Intermittency in an inverse cascade was diagnosed at high order statistics in \cite{jun_05} in a soap-film experiment. A balance between linear friction $\sim \alpha {\bf u}$ and advection ${\bf u} \cdot\nabla {\bf u}$ gives $E(k)\sim k^{-3}$ but possibly in the absence of a constant flux of energy since friction acts at all scales in the inverse cascade range. Moreover, 
removing the large-scale large-intensity vortices leads to a recovery in the inverse range of a $k^{-5/3}$ law again, linking clearly the change in spectral slope to the presence or {absence} of coherent structures. The origin of these coherent structures has been studied in the context of relaxation processes for long times, and of entropy principles (see, e.g., \cite{carnevale, mcwilliams, sinhPoisson, servidio}, and also also \cite{Schorghofer} for surface QG flows, and the case of a passive tracer in such a 2D turbulence).

\section{Inverse cascades}
\subsection{Forcing and dissipation mechanisms and lack of universality in 2D flows}

In a series of experiments it was found that, depending on the bottom-wall friction, two different regimes could be observed: for weak forcing, a condensate forms {via the} merging of smaller vortices, with a steep $\sim k^{-3}$ spectrum, whereas for strong forcing, a $k^{-5/3}$ cascade is observed with clustering of vortices at a size comparable to that of the forcing scale  \cite{paret_98}. The influence of the large-scale drag on the inverse cascade and on structure formation was also investigated in \cite{suko}. From the numerical standpoint, the analysis of such spectral laws shows that they depend also on whether or not both the direct and inverse ranges are properly resolved \cite{danilov_01b, scott}; as an example, hyperviscosity and a short enstrophy range inhibits the formation of the large-scale vortex and thus favors a $k^{-5/3}$ law. Furthermore, the disruption of vortices into filaments can only happen when the small-scale range (to which the filaments belong) 
 is sufficiently well resolved, in turn affecting the steepness of the resulting energy spectrum \cite{vallgren_2011a}. There are also some indications that, as the Reynolds number is increased sufficiently so that an enstrophy cascade develops, the spectrum of the inverse energy cascade steepens, a phenomenon again related to the more efficient formation of large-scale coherent structures \cite{scott}.
 
Some degree of non-universality in the structures and statistics of the flow has thus been observed, due to several factors: the nature of the forcing term, the friction (or hypo-viscosity) used at large scale to prevent an accumulation of energy on the gravest mode, the eventual dissipation mechanism at small scales, as well as boundary conditions and the overall geometry (see, e.g., \cite{chavanis, Elhmaidi, bracco_10}). As an example of the latter factor, it was shown analytically in \cite{chavanis} that a single vortex forms in a square domain, but a dipole forms for rectangular boxes of sufficient aspect ratio (greater than $\approx 1.12$). It was also found in \cite{bracco_10} that the correlation time of the forcing function matters in determining the shape of vortices, although very long time statistics are needed in order to observe {the} effect. 

The observed correlation between small- and large-scale dynamics implies non-locality in Fourier space and non-universality \cite{danilov_01b}, and implies as well that both cascade ranges have to be explicitly incorporated in the flow, with sufficient resolution at small and large scales. Non-local interactions seem to be particularly evident when polymers are added to the flow: they affect the small scales and produce a drag reduction, but are also known to affect the inverse cascade of energy at large scale; this was shown using a shell model \cite{benzi_03} and in experiments \cite{jun_06}. Another example of scale non-locality comes from the recent study in \cite{bracco_10}, where the statistics of vortex population is analyzed. However, it should be noted that a wavelet-based analysis of 2D turbulence was performed in \cite{fischer_09}, concluding that enstrophy transfer is local in configuration space.

In recent years, higher resolutions {have been} achieved in {2D} simulations \cite{vallgren_2011a, bracco_10, Bernard06}, with up to $32768^2$ grid points, leading to sizable cascades, and with a choice of forcing scale of up to roughly 1/1000 the size of the overall computational box. This has allowed for refined statistics, including a detailed study of the conformal invariance properties of the inverse cascade of energy \cite{Bernard06}. However, the presence or not of a logarithmic correction to the small-scale spectra as predicted by Kraichnan \cite{rhk_71}, insuring locality of interactions, is still open to debate. A moderate resolution run with very long time integration (of the order of 1000 turn-over times) and using both linear friction and hyper-viscosity finds such a correction \cite{pasquero},  and it is found as well in \cite{vallgren_2010} in the case of 3D QG turbulence. On the other hand, no such correction appears in the 2D case at substantially higher resolutions (but not necessarily higher Reynolds numbers, given the choice of forcing wavenumber) {or} in the absence of large-scale friction \cite{vallgren_2011a} (see also \cite{avelius}).

\subsection{The case of quasi two-dimensional flows}

Two-dimensional fluid turbulence is thus being explored and reassessed today, in view of several new results stemming mostly (but not uniquely) from direct numerical simulations (DNS), with new findings for systems that are closely related but not identical to the original 2D-2C (2D, two velocity components) case. But how much can one depart from the standard 2D-2C case, and still observe an inverse cascade of energy? Indeed, the experimental flows discussed above are at best quasi-2D, so that one can inquire whether the finite thickness of the fluid with which the experiments are carried out, or the {existence and/or} roughness of the boundary layer play a role in the dynamics of such flows. Flows in nature, such as the atmosphere and the oceans, are also quasi-2D. 

This question has been tackled recently using a variety of approaches.
The inverse cascade of energy in quasi-2D flows is a phenomenon already observed in the framework of shell models \cite{bell, boffetta_shell}. For example, in \cite{bell}, adjusting the parameters in the model allowed for a quasi-2D behavior insofar as a dual (direct/inverse) cascade was observed, and in \cite{chakraborty}, the transition from three-dimensional to two-dimensional behavior was attributed to a preferred transfer to $k_z=0$ Fourier modes when rotation was imposed in the $z$ direction.

Recent DNS using a 2D (say, horizontal) forcing term in a 3D box with a varying aspect ratio in the vertical direction show that, already for an aspect ratio smaller than $1/2$, an inverse cascade of energy begins to develop with a linear growth of the energy \cite{celani} (see also \cite{smith_96}). The growth rate approaches the injection rate of energy as the aspect ratio gets smaller; in fact, it was found in \cite{scott} that the strength of the cascade increases in a monotonic fashion as the Reynolds number (i.e., the extent of the direct enstrophy cascade) grows.  Similarly, in quasi-2D experimental flows with thick layers of fluid, it was shown in \cite{xia_nature} that a strong planar vortex suppresses vertical eddies through vertical shear with an associated time-scale that is shorter than the {vertical} eddy turn-over time {(thus enabling it to
quench vertical motions)}, and that small-scale and large-scale forcing can combine to form an inverse energy cascade. A similar reversal of cascade direction {was found in \cite{verga_passive}} for a passive tracer embedded in a compressible flow.
However, it is not clear whether a supersonic flow, as encountered for example in the interstellar medium, will reach such a degree of compressibility since shocks form and decay rapidly whereas vortices build on an eddy turnover time \cite{kristsuk}.

\begin{figure*}[h!tb] \begin{center}
 	\includegraphics[viewport=43 189 549 586,clip,width=58mm]{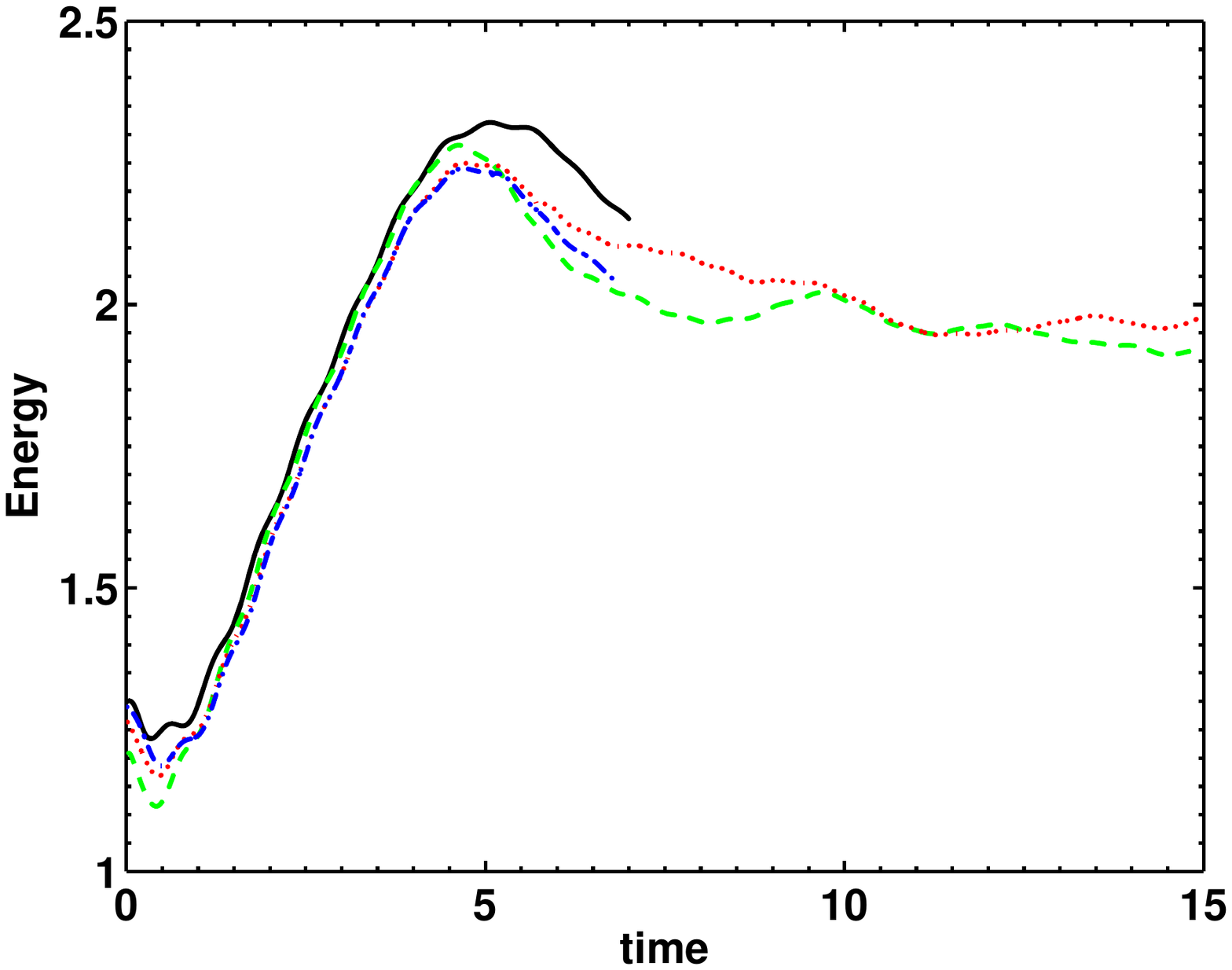}
 	\includegraphics[viewport=35 189 549 586,clip,width=58mm]{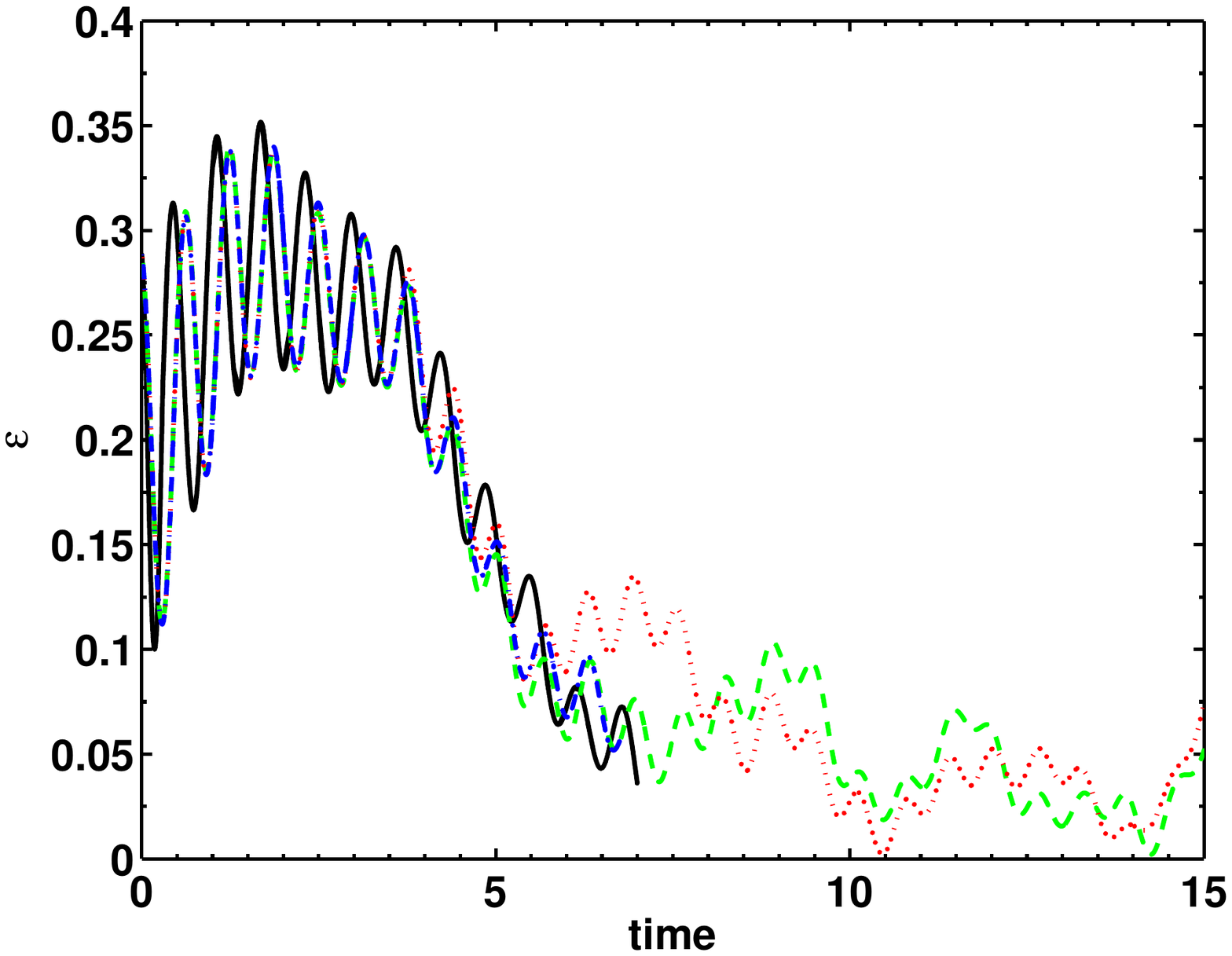}     
 	\includegraphics[viewport=21 184 540 593,clip,width=58mm]{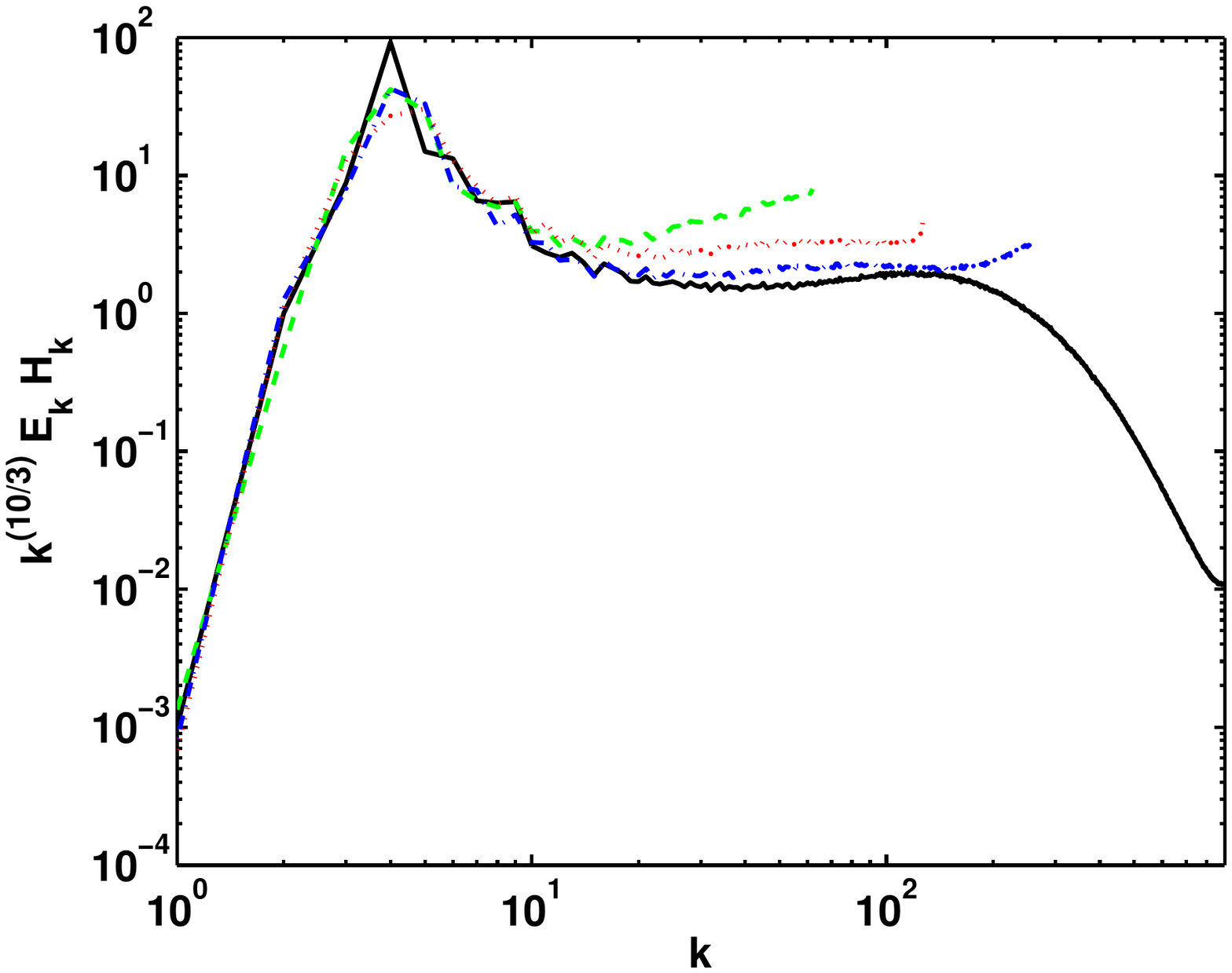} 
 	\end{center}
 \vskip-0.3truein
	\caption{
Temporal evolution of the total energy $E$ (left) and injection rate $\epsilon$ (middle) for the DNS (black) and for several LES: $128^3$ (green dash), $256^3$ (red dots), and $512^3$ (blue dash-dot). The Zeman wavenumber can be recovered using $\Omega=5$. At right  is the product of the energy and helicity spectra compensated by $k^{10/3}$ for the same runs and averaged for $t\in [4.5, \  6.5]$.
} \label{3072a} \end{figure*}

\subsection{Restricted 3D helical flows}

In a traditional view of {incompressible fluid} turbulence, the invariants of the ideal equations (those in the absence of dissipation) determine to a large extent the dynamics of the dissipative case. If the energy invariance of an ideal 3D flow is well-known, such is not {necessarily} the case for the other quadratic invariant{--helicity--} as observed{, say,}  in the vicinity of tornadoes \cite{colquhoun_96}. Helicity is defined as the correlation between the velocity ${\bf u}$ and its curl, the vorticity ${\mathbf \omega}=\nabla \times {\bf u}$, namely $H=\int {\bf u} \cdot {\mathbf \omega} \ d^3 {\bf x}$; note that helicity is not definite positive, which renders the interpretation of its Fourier dynamics and fluxes more complex (see, e.g., \cite{chen_2003H} for a detailed analysis). A helical flow can be viewed as a superposition of helically ($s=\pm$) polarized waves \cite{craya, herring, waleffe} written as:
$$
{\bf u}({\bf k}) = \Sigma_{s}  a_s({\bf k}){\bf h}_s({\bf k}), \ {\bf h}_{\pm}({\bf k})= \hat p \times \hat k \pm i \hat p\ , \  {\bf p}({\bf k})=\hat z \times {\bf k} \ ,
$$
with the hat defining vectors of unit length and $z$ being an arbitrary direction, conveniently chosen in the rotating case as the axis of rotation, making the ${\bf h}_{\pm}$ functions helically polarized inertial waves. This helical wave decomposition has recently been generalized to the case of channel flows, including with streamwise rotation \cite{yang}. Note that there are four types of basic triadic interactions {for wavevectors ${\bf k}$, ${\bf p}$, ${\bf q}$, defined by the helicity modes $(s_k,s_q,s_q)$}, namely {$(s_k,s_q,s_q)= $} (+++), (++$-$), (+$-$+) and (+$- -$), plus four more exchanging the two ($+$  and $-$) polarities \cite{waleffe}. 

For 3D turbulence in the presence of helicity, only a dual energy and helicity cascade to small scales has been found in numerous DNS, with both following a Kolmogorov $k^{-5/3}$ law. Loosely speaking, being more of a small-scale field since it weighs more the small scales than does the energy, the helicity will undergo a direct cascade to small scale more readily than the energy does. However, and perhaps more surprisingly, an inverse cascade of energy (with a direct cascade of helicity) was observed recently  for 3D helical turbulence when restricting the nonlinear interactions to the subsets of Fourier modes that have only one-sign (either $+$ or $-$) polarity \cite{biferale_hel}. In the general case waves of different signs can interact  (see \cite{waleffe}), but since individual triadic nonlinear interactions conserve the invariants separately, one can indeed truncate the equations to such subsets of modes. It is interesting to point out that when dealing with all interactions ($+$ and $-$) the helicity cascade is difficult to study as its flux may also change signs. Restricting interactions to same-sign modes does not have such impediment, and a direct cascade of helicity can be identified clearly. The invariance of one-signed helicity in this restricted case leads to the inverse cascade of energy to the large scales which follows again a $k^{-5/3}$ spectrum \cite{biferale_hel}.

This result does not contradict the previous numerical findings using the full set of nonlinear interactions; indeed, it was shown in \cite{rhk_73} that same-signed interactions are sub-dominant and thus one can recover the traditional direct cascade of energy for 3D fluid turbulence in the full case. The recent study in \cite{biferale_hel} does show, however, that the principles of statistical mechanics on the basis of which arguments can be developed in favor of direct and inverse cascades, are sturdy and extend as well, perhaps in unexpected ways, to (carefully selected) subsets of modes. Sub-ensembles of modes having different scaling laws is a known feature of turbulent flows, as for example in Rayleigh-B\'enard convection when differentiating between the (0,0,2n) modes of the temperature and all other modes, due to inherent symmetries of the equations \cite{mishra_11}. The {\it gedanken} numerical  experiment restricting interactions to the $+++$ triads also gives credence to the observation that, when examining the individual energy transfer in triads, there are numerous interactions transferring energy to large scales, some of which can be interpreted as being due to this subset of same-signed helical modal interactions, and others that obviously correspond to purely 2D triads.

\subsection{Decaying versus forced flows}

Inverse cascades are traditionally viewed as being a hallmark of forced turbulence: it is commonly thought that forcing is necessary to observe an inverse cascade, in particular because of the energy needed to populate the large scales, with a linear growth of the total energy {in time}. However, it has been shown recently using an ensemble of numerical simulations computed on grids of $2048^2$ points, that a large-scale $k^{-5/3}$ spectrum can be observed as well in the decaying case at scales larger than the initial conditions, when taking an average over long times and over the ensemble \cite{mininni_2d}. This observed spectrum corresponds to an inverse cascade with a constant negative flux for the ensemble. Such a behavior is not necessarily surprising since it  is contained in the nonlinearity of the primitive equations but it has implications for experimental quasi-two-dimensional flows which are found to follow in some cases a $k^{-5/3}$ law even though the flow is interpreted as being decaying.

\subsection{Break down of 2D effects and rotating flows}

\begin{figure*}[h!tb] \begin{center}   
   \includegraphics[viewport=26 184 553 593,clip,width=58mm]{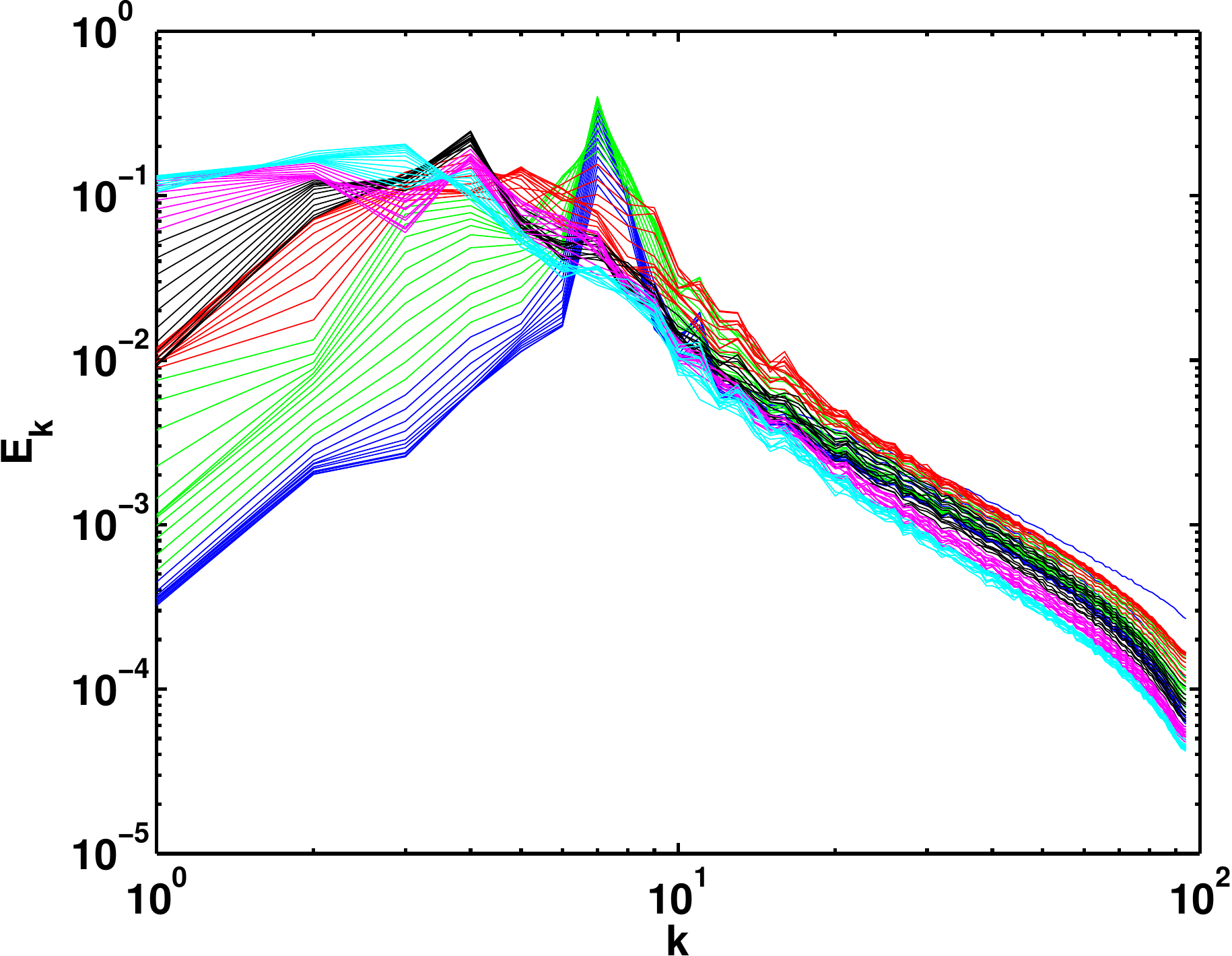} 	
   \includegraphics[viewport=24 184 553 586,clip,width=58mm]{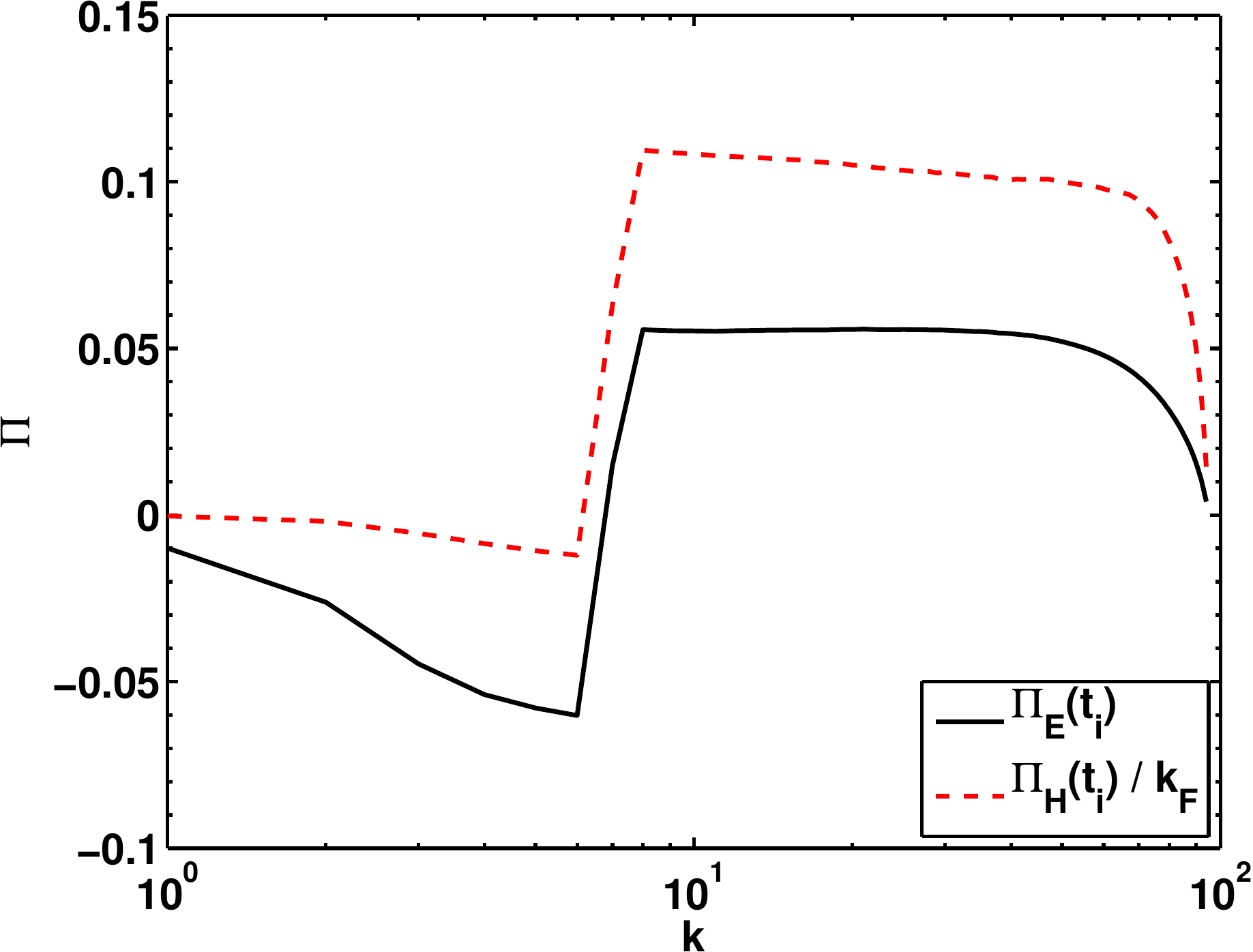} 
   \includegraphics[viewport=24 184 553 586,clip,width=58mm]{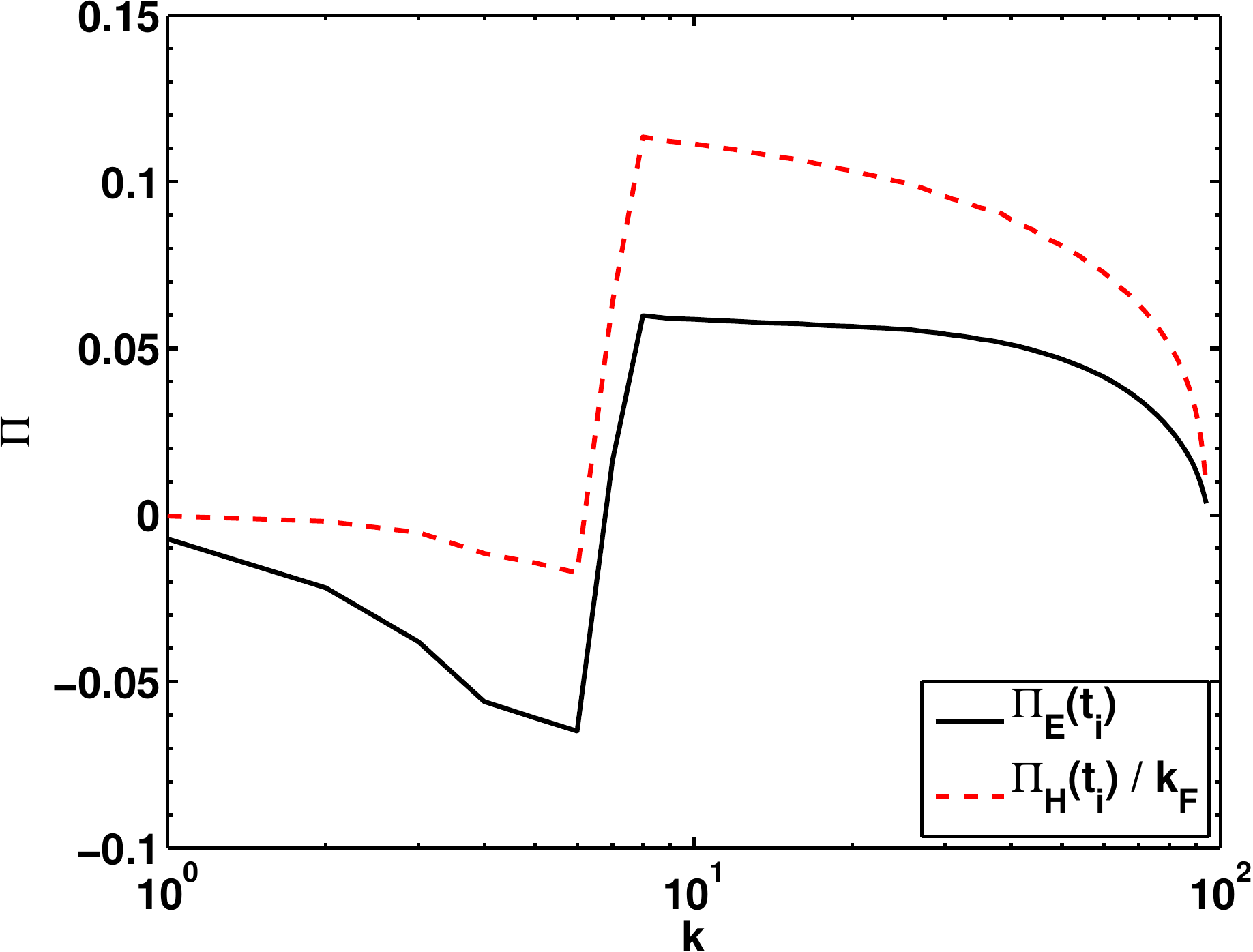}
\end{center}
\vskip-0.3truein 	
 \caption{
{\it Left:} Detailed evolution with time of the energy spectrum for a run with $Re\approx 6.2 \times 10^4$ and $Ro=0.014$. One sees that the effect of rotation is to populate the large scales at the expense of the small scales which are depleted.
 Energy (black) and helicity (red dashed) spectral fluxes at the end of the computation for two runs with high ({\it middle}) and moderate (4 times smaller, {\it {right}}) Reynolds number but with otherwise comparable Rossby numbers and numerical resolutions.
 Constancy of flux of energy and helicity to the small scales only occurs at the higher Reynolds number.
 	} \label{f:julien} \end{figure*}

The reverse effect of the selection of 2D modes in initially quasi-2D flows, or in other words, the three-dimensionalization of 2D turbulence, has also been observed, e.g., in freely decaying turbulence \cite{ngan}. Similarly, forced turbulence with solid body rotation $\Omega$, which behaves in a quasi-2D way at large scale, has been shown recently to recover its 3D properties at small scales \cite{mininni_3072}, provided the so-called Zeman scale $\ell_{\Omega}=[\epsilon/\Omega^3]^{1/2}$ with $\epsilon=\dot E$
(equivalent to the Ozmidov scale for stratified flows) is resolved (the Zeman and Ozmidov scales are defined as the scales at which the turnover time and the wave period are equal). One may have to distinguish here between decaying flows, for which the Reynolds and Rossby numbers vary in time in different ways, and the forced case in which these two dimensionless parameters can be kept relatively constant. In the decaying case, the ratio of  Zeman  to dissipation scale varies with time (the Zeman scale getting smaller as the Rossby number gets smaller), whereas in the forced case this ratio does not change substantially and could even be constrained to retain a constant value. Thus, the fact that anisotropy is found to be stronger at smaller scales in a laboratory experiment analyzed recently \cite{moisy11} may be related to the  variation of  this ratio of scales and is not necessarily in contradiction with DNS results of isotropization of the small scales in the presence of forcing (although non-monotonic effects in rotating flows cannot be discarded either). Again, universality should not be assumed too rapidly when several phenomena with different time scales are in competition, and more experiments, {both numerical and in the laboratory}, are clearly needed.
	
What remains of these findings in the presence of imposed solid body rotation? Of course, the computational demand is substantially greater than in 2D, and it will take time to explore parameter space {thoroughly}, but results are already appearing that emphasize the role played by the nature of the forcing (emphasizing or not 2D versus 3D modes) at large scales \cite{smith_99}. It is in this context that {we wish to address briefly the extent to which sub-grid scale models of turbulence can be used to further explore parameter space in rotating turbulence. To this end}, we {present here} new tests of a previously developed Large Eddy Simulation (LES) model of turbulence \cite{julien}, {for which the Zeman scale may or may not be resolved} {against a rotating turbulence DNS performed on a grid of $3072^3$ points, for which the Zeman scale
is well resolved.} We then mention recent results on the inverse cascade of energy in rotating 3D turbulence using this model, and we end with 
some concluding remarks.

\section{Resolving or not the Zeman scale in a model of rotating turbulence}

Let us begin by writing the Navier-Stokes equations in a rotating frame of reference:
\begin{equation}
\frac{\partial {\bf u}}{\partial t} + \mbox{\boldmath $\omega$} 
    \times {\bf u} + 2 \mbox{\boldmath $\Omega$} \times {\bf u} =
    - \nabla {\cal P} + \nu \nabla^2 {\bf u}  + {\bf F } \ ; 
\label{eq:momentum}
\end{equation}
${\cal P}$ is the total pressure  modified by the centrifugal term; it is obtained by taking the divergence of Eq.~(\ref{eq:momentum}) and assuming incompressibilty, $\nabla \cdot {\bf u}=0$. The rotation $\mbox{\boldmath $\Omega$}$ is imposed in the vertical ($z$) direction. The Reynolds number  is $Re=U_0L_F/\nu$ with $U_0$ the r.m.s.~velocity, $L_F=2\pi/k_F$ the forcing scale, and $\nu$ the kinematic viscosity, and the Rossby number is $Ro= U_0/(2 L_F \Omega)$; ${\bf F}$ is a forcing term.

As stated before, together with energy, helicity is a global invariant of the Euler equations (see \cite{moffatt_tsinober} for  a review) including in the presence of solid body rotation. In Fourier space,  the relative rate of helicity is $\rho(k)=H(k)/[2kE(k)]$ with $H= \int H(k)dk$ and $E= \int E(k)dk$ the total energy. A Schwarz inequality implies that $|\rho(k)| \le 1$, and the helicity is said to be maximal when $\rho(k)=\pm 1$. It can be created locally in space through the correlation of vorticity and shear or pressure gradients \cite{matthaeus}, and globally through a combination of stratification and boundaries or rotation \cite{moffatt_tsinober}. Fully helical {flows} (with aligned velocity and vorticity everywhere) are called Beltrami, and can be represented by the so-called ABC flow:
\begin{eqnarray}
{\bf u} / u_0=&[B cos(k_0y) + C sin(k_0z)]\hat x  \\
+ &[C cos(k_0z)
+ A sin(k_0x)]\hat y \\
+ &[A cos(k_0x) + B sin(k_0y)] \hat z \ .
\end{eqnarray} 
Such flows are not attractors of statistical ensembles in ideal fluids \cite{rhk_73} and are not globally stable \cite{podvigina}, but, in helical turbulent flows in the absence of rotation, full mirror-symmetry (no helicity) recovers slowly with wavenumber, as $1/k$, since both the energy and helicity have Kolmogorov $k^{-5/3}$ spectra. Note that the ABC flow is an intrinsically 3D flow; it excites only a few modes selected by its helical symmetry in a given shell (basically along the three $k_x, \ k_y, \ k_z$ axes), with dependence in all three directions, and it has three components of the velocity field. However, the amount of energy in 2D vs. 3D modes remains constant as one increases $k_0$. On the other hand, for isotropic initial conditions or forcing, the amount of energy in 2D vs. 3D modes decreases as one increases the wavenumber of the flow since the energy in the 2D modes grows as $k_0$, while the energy in 3D modes grows as $ k_0^2$, i.e., as the number of points in a spherical Fourier shell. When the ABC flow is concentrated in  the large scales, the resulting anisotropy of the flow is not significant, but for problems of energy decay \cite{teitelbaum_11}, or for inverse cascades for which ABC is used as a forcing and is moved to small scales with forcing wavenumber $k_F=k_0$ of the order of a few tens, the anisotropy becomes measurable and can affect the resulting dynamics \cite{paper1}.

 \begin{figure*}[h!tb]         \begin{center}
        \includegraphics[viewport=26 155 558 602,clip,width=58mm]{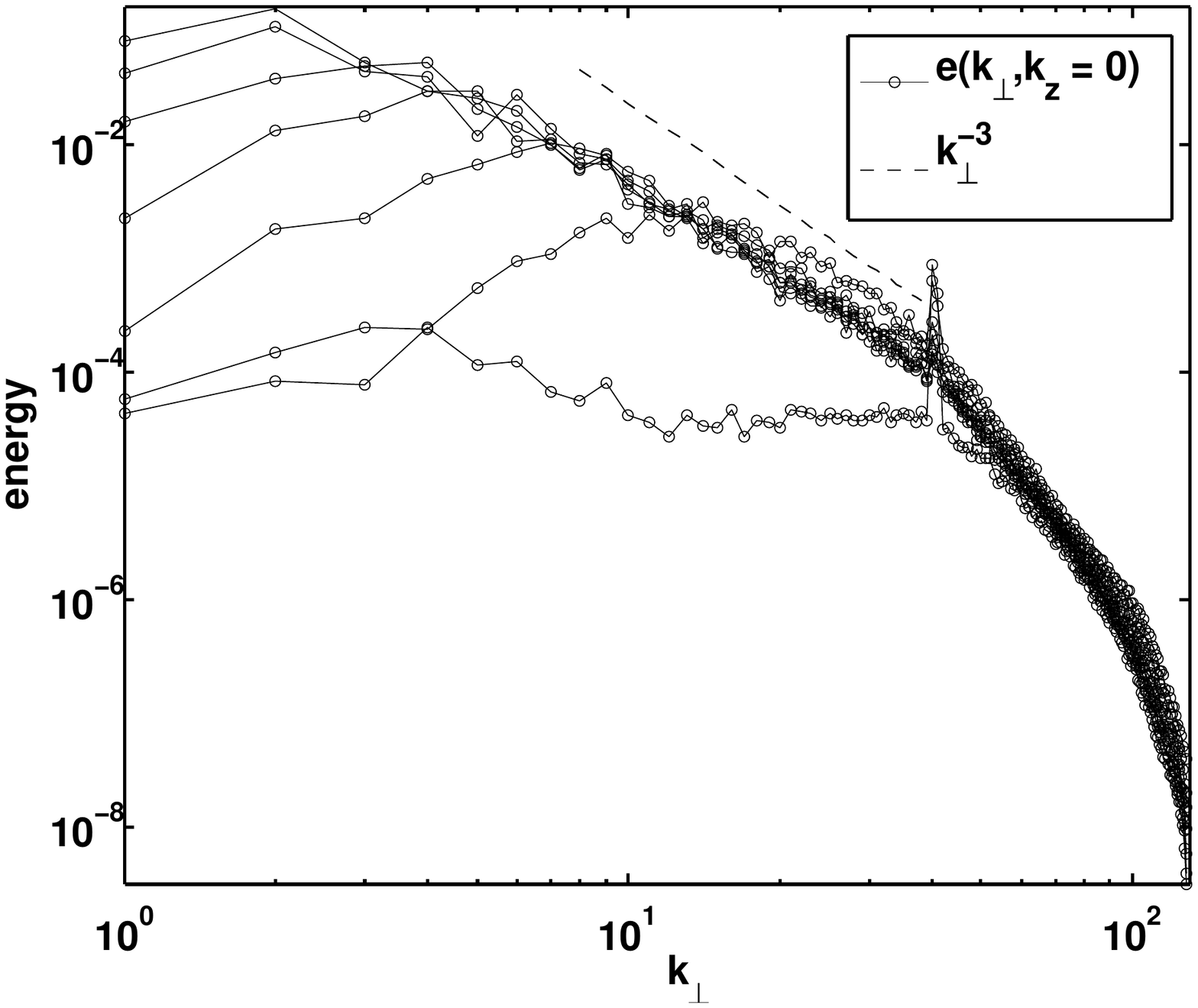}
        \includegraphics[viewport=14 155 557 610,clip,width=58mm ]{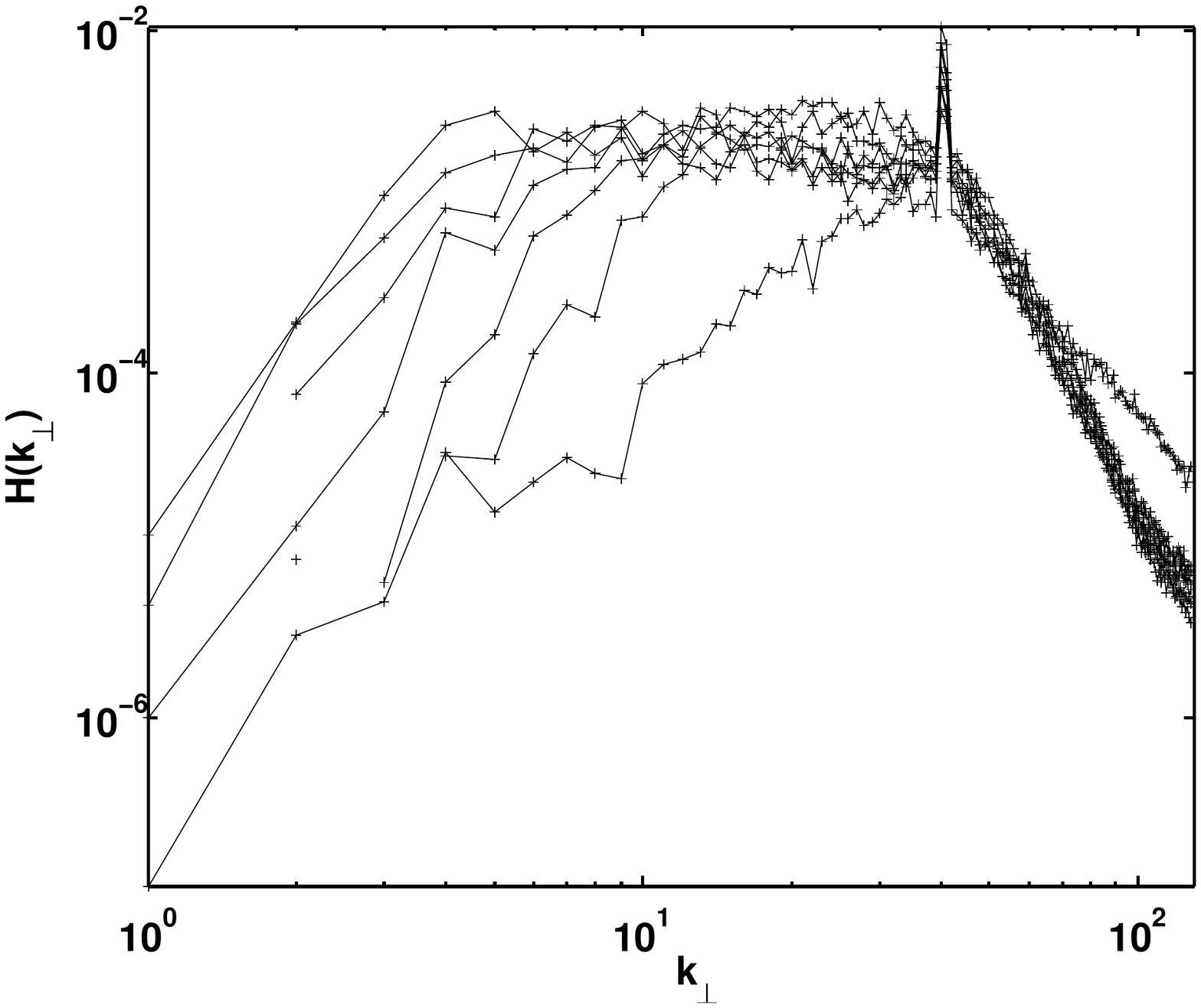}
        \includegraphics[viewport=21 164 557 604,clip,width=58mm]{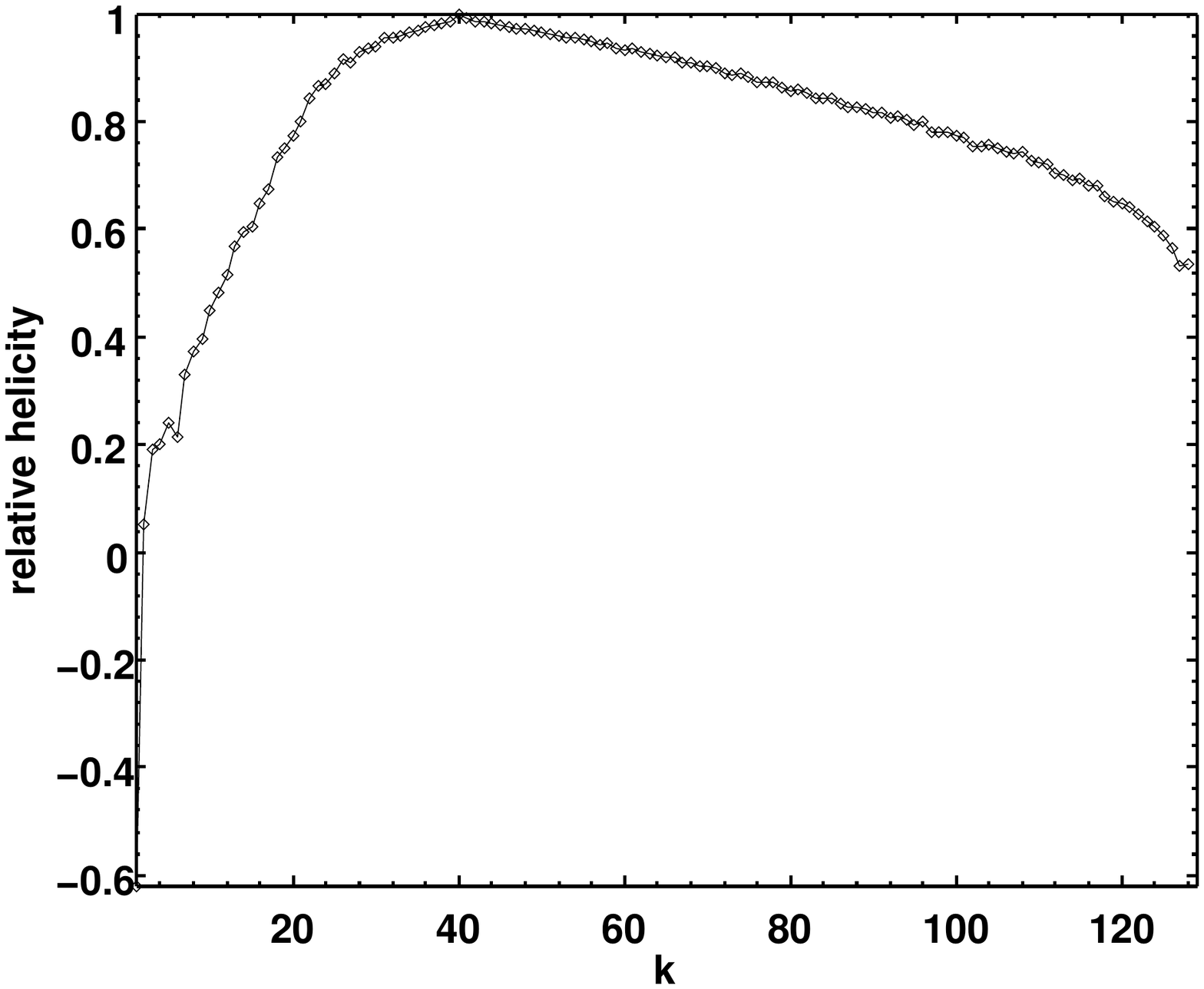}
    \end{center}
 \vskip-0.3truein
\caption{
{\it Left :} Two-dimensional energy spectrum $e_{2D}(k_{\perp}, k_z=0)$ for various times for a random forcing 
with no helicity and $\beta=1$); note the build-up towards a $k^{-3}$ law.
{\it Middle:} Two-dimensional helicity spectrum $h(k_{\perp}, k_z=0)$ for various times for a $\beta=3$  maximal helicity forcing. 
{\it  Right:} Spectrum of relative helicity for the same run.  
} \label{ener_B1r0} \end{figure*}

The DNS of forced rotating turbulence used as ``ground truth'' to validate LES results corresponds to a simulation with $3072^3$ points, with parameters $Ro=0.07$ and $Re=2.7 \times 10^4$, chosen such that the Zeman scale is well resolved ($k_{\Omega}\approx 30$). ABC was used as the forcing ${\bf F}$ to inject both energy and helicity, and applied at $k_F=k_0=4$. The LES we employ is described in detail in \cite{julien}; it has been tested against a DNS on a grid of $1536^3$ points with forcing at $k_F=7$ that has the Zeman scale close to the dissipation scale. The LES model implements both an eddy viscosity and an eddy noise, and these two transport coefficients have both a non-helical and a helical contribution. The eddy viscosity does not assume a Kolmogorov spectrum but rather fits the spectrum at a given time with a power law followed by an exponential decay; the fit is done in the interval $[k_c/3, k_c]$ where $k_c$ is the cut-off wavenumber. Hence the question arises as to whether the LES can capture the behavior of the Zeman scale when this scale lies in the middle of the inertial range. We thus look at several LES runs computed on grids of $128^3$ to $512^3$ points. All flows start from the same initial conditions, a statistically steady state of a turbulent flow without rotation, and thus have initially Kolmogorov spectra both for the energy and the helicity.

Figure \ref{3072a} {shows} the temporal evolution of the total kinetic energy and of the energy injection rate for the DNS and for the three LES runs which are identical except for the resolution of the grid used in the LES. {For none of these runs is a} global energy increase discernible (except for the initial  phase adjustment to rotation), since the inverse cascade of energy is barely resolved, with a forcing at $k_F=4$. The initial growth, up to $t\approx 4.5$ is remarkably similar for all runs, {but} differences become discernible {afterwards}. All LES runs underestimate the energy, by roughly up to 15 \%, and over-estimate the helicity (not shown); 
 Neither the injection rate nor the dissipation differ significantly from run to run {(see Fig. \ref{3072a})}, except for a phase difference in the oscillations, and thus the Zeman wavenumbers computed for {all runs are quite close.} These oscillations linger over time, a phenomenon likely related with the excitation of waves when rotation is turned on \cite{mininni_1536a}, but note that all LES runs {exhibit a delay with respect to the DNS that persists.}  It would be of interest to identify the origin of this discrepancy. 
 
The phenomenological prediction for turbulence in the presence of rotation is that $E(k)\sim k^{-e}$, $H(k)\sim k^{-h}$, with $e+h=4$ at scales where rotation prevails, and $E(k)\sim H(k)\sim k^{-5/3}$ beyond the Zeman scale when isotropy is recovered. In that light we show $k^{10/3}E(k)H(k)$ in the right of Fig. \ref{3072a}.  All spectra are in overall good agreement until the Zeman wavenumber $\approx 30$, at which there is a break in the spectra and they evolve towards a dual Kolmogorov cascade $E(k)\sim k^{-5/3}$, $H(k)\sim k^{-5/3}$ in the DNS \cite{mininni_3072}. This break in the spectra is visible for all LES runs, but clearly too accentuated for the lower resolution LES. The Kolmogorov spectrum for $E(k)$ and $H(k)$ is well recovered, whereas the steeper spectrum in the DNS is {not accurate} except at the highest resolution LES, although the transition between the two is well marked {in all runs}. The fine-tuning between waves and nonlinear steepening due to advection may need an even higher resolution in the LES, in part because the fit to the spectrum catches the Kolmogorov part but perhaps misses, in the evaluation of transport coefficients, the large-scale steeper part.
Finally, note that the error in the global energy has to be put in perspective, realizing the enormous savings in computation time, $\approx (3172/512)^4 > 1400$ for the $512^3$ LES (and more so for the lower-resolution LES). 

In view of these results, as with those given in \cite{julien} for tests of the same LES but against a DNS at lower resolution and not resolving the Zeman scale, we can conclude that  the LES model can reproduce well the dynamics of the anisotropic large-scales even in complex flows with different scaling behavior at different scales. Using this LES, we now show in Fig. \ref{f:julien}  the energy spectrum at different times for a run with $Re\approx 6.2 \times 10^4$ and $Ro=0.014$, forced at $k_F=7$ on a grid of $192^3$ points with the  initial conditions {are that} of a fully-developed turbulent flow without rotation. As time evolves, the small-scale range becomes steeper, and the inverse cascade builds up progressively towards  scales larger than the forcing scale. At the final time, the energy has reached the largest available scale (no large-scale friction is used). The energy and helicity fluxes {are also shown for the same run in
Fig. \ref{f:julien}}, the latter normalized by the forcing scale. We observe, as already mentioned in \cite{mininni_1536a, mininni_3072}, a dominance of the helicity flux to small scales, and of the energy flux of opposite sign to large scales, the helicity flux being negligible for scales larger than $L_F$. Lastly, at right in Fig. \ref{f:julien} are these fluxes for a run with the same Rossby number but a Reynolds number four times smaller,  {in which case the constancy} of the fluxes is not as well realized.

\section{The inverse cascade of energy in rotating turbulence }

When solid body rotation is imposed on the flow, the interactions between inertial waves and nonlinear eddies lead to a slowing-down of the energy transfer to small scales and to a steeper spectrum that can be modeled using a simple phenomenological argument which incorporates the interactions between eddies and waves \cite{dubrulle, zhou}. Furthermore, when helical forcing is used, another scaling law {arises} because of the dominance of the helicity cascade to small scales \cite{mininni_1536a, phil_trans, mininni_3072}. Moreover, the direct cascade of energy is found to be scale-invariant, and in fact conformal invariant \cite{simon}, a stronger local property involving transformations that preserve angles, and which again connects this problem with other quasi-2D flows.

The inverse energy cascade to scales larger than the forcing scale in rotating turbulence has been studied using numerical simulations with  hyper-viscosity in \cite{smith_96} where it was shown that it coexists with the direct cascade of energy. It was also  found that most of the energy resides in the 2D modes, i.e., those with $k_z=0$ \cite{smith_99}.  A further exploration using so-called reduced models  at moderate Rossby number $\sim 10^{-1}$ showed that only the models that include near-resonances reproduce well the (moderately resolved) DNS flows \cite{smith_05}.

New computations of the inverse cascade of energy in 3D rotating flows, both for non-helical and for helical forcing, and for 2D versus 3D forcing, are being conducted presently \cite{paper1}; they use the spectral LES tested in the preceding section, at resolutions of $256^3$ grid points, and some DNS are in the planning stage as well. These investigations lead to the conclusion that the way the forcing occurs, putting more or less weight on 2D modes (i.e., forcing in the horizontal plane) versus fully 3D modes, is the main parameter of the problem and breaks the universality, as already found in \cite{smith_05} when only a subset of the triadic interactions were considered. 

{Given} a random isotropic forcing function ${\bf f}_{\rm RND}({\bf k})$ centered in a narrow band around a given wavenumber $k_f$ and with a certain relative helicity, one can prescribe {in simulations} the amount of 2D versus 3D forcing by defining a new forcing
$$
{\bf f}_{\rm ANI}({\bf k}) = \left(1-\frac{k_z}{k_f}\right)^\beta {\bf f}_{\rm RND}({\bf k}) \ ,$$
with the parameter $\beta$ controlling how much energy is injected in 2D modes, $\beta=0$ corresponding to isotropic forcing. One can then study the scaling of the energy in 2D and 3D modes by decomposing $E$ into its 3D component, $E_{3D}$, with $k_z\not= 0$, and its 2D components for ${\bf u}_{\perp}$ (of energy $E_{\perp}$), and $u_z$ (of energy $E_w$), both with $k_z=0$, i.e., for the subset of modes with no variation in the direction of the rotation, with $E_{2D}=E_{\perp}+E_w$.

Figure \ref{ener_B1r0} shows the evolution of the 2D energy spectrum for a random flow with $\beta=1$ (i.e., for slightly anisotropic forcing) and with $k_F=40$. 
It grows with time and dominates the 3D component at all times except in the onset phase, 
and has almost reached the gravest mode at the final time of the computation, following a $k_\perp^{-3}$ law. This result is also found for several other runs, {except when the forcing is strongly anisotropic (e.g., for the ABC forcing, or for a random forcing with $\beta=3$); in these latter cases, a $k_{\perp}^{-5/3}$ {is obtained} in the inverse cascade of 2D energy \cite{paper1}.} In Fig.  \ref{ener_B1r0} (middle), we show the relative helicity for a helical run as a function of wavenumber. It  is maximal (by construction) at the forcing wavenumber and appears to decrease slowly in the direct cascade, and sharply in the inverse cascade of energy: helicity does not follow efficiently the energy to large scales, as was already clear when plotting the fluxes of helicity and energy (see Fig. \ref{f:julien} and \cite{mininni_1536a}). This is also noticeable when examining
the spectrum of {2D} helicity  given at far right in Fig.~\ref{ener_B1r0} for several times as well during the inverse cascade phase for the energy; in theses units, the spectrum is flat, indicating that no cascade to large scale is taking place for the helicity, corroborating the observation made when looking at the flux of helicity.

We finally note that the dynamics resulting from an ABC forcing behaves in a unique way: in this case only, the 2D modes still follow a $\sim k_\perp^{-5/3}$ spectrum; however, the {isotropic} energy spectrum in the inverse cascade range follows a $k^{-1}$ law, due to strong anisotropy in the forcing and with a strong influence of shear at large scales \cite{paper1}. 

\section{Conclusions}
	
Quasi-two dimensional flows show a richness of behavior and the detailed properties of the inverse cascade to large scales can vary according to the {flow} set-up {considered}. It is likely that such a diversity will be unraveled as well for three-dimensional flows in the presence of rotation, as already shown in \cite{smith_05} using a model that preserves a subset of triadic interactions. In the present paper, we first tested a previously developed sub-grid scale model  of turbulence against  a large direct numerical simulation performed on a grid of $3072^3$ points; we showed that the transition from a large-scale anisotropic to a small-scale isotropic inertial range is well reproduced granted the LES is sufficiently resolved; with such models, one can study parameter space at a reasonable computational cost. We then used the model to confirm that for rotating turbulence, the inverse cascade of  energy can have two different scaling laws according to how anisotropic the forcing is \cite{paper1}, with moreover shear playing an important role for strongly anisotropic forcing when considering the ABC flow. It remains to be seen that other strongly anisotropic flows behave in a similar fashion; this could be examined for example by taking a higher value of the parameter $\beta$ in the random forcing function introduced in the preceding Section.
The non-uniqueness of the scaling in the inverse cascade of rotating turbulence, both for the 2D and the 3D spectra, clearly needs further investigation. So does the 
singularity of behavior for the ABC forcing, leading to a $k^{-1}$ scaling for the total energy: depending on the extent of anisotropy in the external forcing function, shear may be introduced in the flow thereby altering the nature of the energy exchange between scales. In this context, it is noteworthy that helicity and strong shear can be strongly coupled, as observed in the atmosphere, for example in the context of tornado and hurricane dynamics in the vicinity of deep convective cells \cite{molinari, montgomery_levina}.

{\it This work is sponsored by an NSF cooperative agreement of NCAR; computer time was provided by NSF TeraGrid projects ASC090050, TG-PHY100029 \& 1025183.}

 \end{document}